\documentstyle[aps,epsf,rotate,preprint,tighten]{revtex}

\begin{document}

\draft

\title{Scaling behavior in economics: II. Modeling of company growth}

\author{Sergey~V. Buldyrev, $^1$ Lu\'{\i}s~A.~Nunes Amaral,$^{1,2}$
Shlomo Havlin,$^{1,3}$\\ Heiko Leschhorn,$^{1}$\protect\footnote{Present
Address: Theor. Physik III, Heinrich-Heine-Univ., D-40225 
D\"usseldorf, Germany.}  Philipp
Maass,$^{1}$\protect\footnote{Present Address: Fakult\"at f\"ur Physik,
Universit\"at Konstanz, D-78434 Konstanz, Germany.}
Michael~A.~Salinger,$^4$ H.~Eugene Stanley,$^1$\\ and
Michael~H.~R.~Stanley$^{1}$\protect\footnote{Present Address: Department
of Physics, MIT, Cambridge, MA 02139.}}

\address{$^1$Center of Polymer Studies and Department of Physics,\\
                Boston University, Boston, MA 02215, USA\\
         $^2$Institut f\"ur Festk\"orperforschung, Forschungszentrum
                J\"ulich, D-52425 J\"ulich, Germany\\
         $^3$Department of Physics, Bar-Ilan University, Ramat Gan,
                Israel\\
         $^4$School of Management, Boston University, Boston, 
                MA 02215, USA
}

\date{last revised: October 31, 1996; printed: \today}

\maketitle

\begin{abstract}

  In the preceding paper we presented empirical results describing the
  growth of publicly-traded United States manufacturing firms within the
  years 1974--1993. Our results suggest that the data can be described
  by a scaling approach. Here, we propose models that may lead to some
  insight into these phenomena.  First, we study a model in which the
  growth rate of a company is affected by a tendency to retain an
  ``optimal'' size.  That model leads to an exponential distribution of
  the logarithm of the growth rate in agreement with the empirical results.
  Then, we study a hierarchical tree-like model of a company that
  enables us to relate the two parameters of the model to the exponent
  $\beta$, which describes the dependence of the standard deviation of
  the distribution of growth rates on size. We find that $\beta = -\ln
  \Pi / \ln z$, where $z$ defines the mean branching ratio of the
  hierarchical tree and $\Pi$ is the probability that the lower levels
  follow the policy of higher levels in the hierarchy.  We also study
  the distribution of growth rates of this hierarchical model.  We find
  that the distribution is consistent with the exponential form found
  empirically.

\end{abstract}

\section{Introduction}

The concept of scaling supports much of our current conceptualization on
the general subject of how complex systems formed of interacting
subunits behave.  This concept was developed a quarter century
ago by physicists interested in the behavior of a system near its
critical point.  Progress was made possible by a remarkable combination
of experiment and phenomenological theory. In the preceding paper we
presented empirical results suggesting that the scaling concept can be
useful in describing economic systems
\cite{Mandelbrot63x,Mantegna95x}. In this paper we present models which
may lead to an understanding of the underlying mechanism behind the
scaling laws.

In the preceding paper, we used the Compustat database to study all
United States (US) manufacturing publicly-traded firms from 1974 to
1993.  The Compustat database contains $20$ years of data on all
publicly-traded companies in the US.  We found that the distribution
of firm sizes remains stable for the 20 years we study, i.e., the mean
value and standard deviation remain approximately constant.  We
studied the distribution of sizes of the ``new'' companies in each
year and found it to be well approximated by a log-normal. However, we
find (i) the distribution of the logarithm of the growth rates, for a
growth period of one year, and for companies with approximately the
same size $S_0$ displays an exponential form \cite{Stanley2,Empirical}
\begin{equation}
p(r_1|S_0) = \frac{1}{\sqrt{2}\sigma_1(S_0)} \exp \left(
-{\frac{\sqrt{2}~|r_1-\bar r_1|}{\sigma_1(S_0)}} \right)\,,
\label{e-distribution}
\end{equation}
and (ii) the fluctuations in
the growth rates --- measured by the width of this distribution
$\sigma_1$ --- scale as a power law \cite{Empirical}, 
\begin{equation}
\sigma_1(S_0) \sim {S_0}^{-\beta}\,.
\label{e-sigma}
\end{equation}
Here $r_1=\ln(S_1/S_0)$, where $S_1$ is the size of the company in the
next year, and $\sigma_1(S_0)$ is the standard deviation (width) of the
distribution (\ref{e-distribution}).  
We found that the exponent $\beta$ takes the same value, within the
error bars, for several measures of the size of a company.  In
particular, we obtained: $\beta=0.20\pm0.03$ for ``sales.''

In this paper, we present and discuss models that, although very
simple, give some insight into these empirical results.  The paper is
organized as follows.  In Sect.  II, we discuss a model that predicts
an exponential distribution of growth rates.  In Sect.  III, we study
a hierarchical tree model that predicts the power law dependence of
$\sigma_1$ on size.  In Sect. IV, we discuss how the two models can be
combined so that a single model predicts both of our central empirical
findings.  In Sect. V, we summarize our findings and suggest avenues
for future research.  The paper contains three appendices. Appendix A
discusses the relationship between the standard deviations of the
growth rate and the logarithmic growth rate. Appendices B and C give
more details of the analytical solution of the hierarchical tree
model.

\section{The exponential distribution of growth rates}

As described above, one of our central findings is that the
distribution of growth rates for companies of a given initial size has
an exponential form.  The result is surprising because the sales of
organizations as large as publicly traded corporations reflect a large
number of factors.  While those factors are not necessarily
independent and while the growth of any one company might be dominated
by a single factor, one might nonetheless expect a Gaussian
distribution for growth rates.

In this section, we show how a plausible modification of
Gibrat's assumptions \cite{Gibrat} could lead to Eq. (1).  We relax
the assumption of uncorrelated growth rates and assume that the
successive growth rates are correlated in such a way that the size of
a company is ``attracted'' to an optimal size $S^*$.  This value is
reminiscent of the minimum point of a ``U-shaped'' average cost curve
in conventional economic theory and should evolve only slowly in time
(on the scale of years) \cite{Samuelson}.  

Let us then consider a set of companies all having initial sales
$S_0$. As time passes, the sales of each of the firms varies from day
to day (or over another time interval much less than 1 year), but
tend to stay in the neighborhood of $S^*$.  In the simplest case, the
growth process has a constant ``back-drift,'' i.e.
\begin{equation}
\frac{S_{t+\Delta t}}{S_t} = \cases{ k(1+\epsilon_t), & $S_t<S^*$, \cr
                      \frac{1}{k}(1+\epsilon_t), & $S_t>S^*$, \cr }
\label{model1}
\end{equation}
where $k$ is a constant larger than one and $\epsilon_t$ is an uncorrelated
Gaussian random number with zero mean and variance $\sigma_\epsilon^2\ll
1$. These dynamics are similar to what is known in economics as
regression towards the mean \cite{Leonard,Friedman}, although this
formulation is not standard in economics.

Written in terms of the logarithmic growth rate $r_t \equiv
\ln(S_t/S_0)$, Eq.~(\ref{model1}) reads
\begin{equation}
r_{t+\Delta t} - r_t = -\ln k~{\rm sgn}(r_t-r^*)
+ \ln (1 + \epsilon_t),
\label{model1_r}
\end{equation}
where $r^* \equiv \ln(S^*/S_0)$ and ${\rm sgn}~x = -1$ for $x<0$ and
${\rm sgn}~x = 1$ for $x>0$.  Since $\sigma_\epsilon\ll1$, we can write
$\ln(1+\epsilon_t)\simeq\epsilon_t$.

For large times $t\gg\Delta t$ we can replace Eq.~(\ref{model1_r}) by
its continuum limit and obtain
\begin{equation}
\Delta t \frac{dr(t)}{dt} = -\ln k{d\over dr}|r(t)-r^*| +
\sqrt{\Delta t} ~ \epsilon(t)\,,
\label{contlimit}
\end{equation}
where now $\epsilon(t)$ is a Gaussian random field with $\langle
\epsilon(t) \rangle=0$ and $\langle \epsilon(t) \epsilon(t') \rangle =
\sigma_\epsilon^2 \delta(t-t')$ \cite{Note1}.  Here, $\langle \cdots
\rangle$ means an average over realizations of the disorder and $\delta$
is the Dirac delta function.
Equation~(\ref{model1_r}) describes a strongly overdamped Brownian
motion of a classical particle with mass one in a potential
\begin{equation}
V(r) = \ln k~|r-r^*|,
\end{equation}
where the friction constant is $\Delta t$ and the thermal energy is
$\sigma_\epsilon^2/2$ \cite{Risken}.  For large times $t\gg\Delta t$
(e.g., after one year), the ``particle coordinate'' $r$ is distributed
according to the equilibrium Boltzmann distribution,
\begin{equation}
p(r_1|s_0)=\frac{\ln k}{\sigma_\epsilon^2}
\exp \left( -\frac{2 \ln k~|r_1-r^*|}{\sigma_\epsilon^2} \right).
\label{model12}
\end{equation}
Hence, we recover Eq.~(\ref{e-distribution}) with $\bar r(s_0)=r^*$
and
\begin{equation}
\sigma_1(s_0) = \frac{\sigma_\epsilon^2}{\sqrt{2} \ln k}.
\label{model11}
\end{equation}

\section{The scaling exponent $\beta$}

While the model in the previous section explains
Eq.~(\ref{e-distribution}), it does not predict our finding about the the
power law dependence of the standard deviation of growth rates on firm
size.  In this section, we show how a model of management hierarchies
can predict Eq.~(\ref{e-sigma}).  In economics, it is generally
presumed that the growth of firms is determined by changes in demand
and production costs.  Since these features are specific to individual
markets, it is surprising that a law as simple as equation
Eq.~(\ref{e-sigma}) governs the growth rate of firms operating in much
different markets.  While demand and technology vary across markets,
virtually all firms have a hierarchical decision structure.  One
possible explanation for why there is a simple law that governs the
growth rate of all manufacturing firms is that the growth process is
dominated by properties of management hierarchies \cite{Radner}.  This
focus on the technology of management rather then technology of production
as a basis for understanding firm growth is reminiscent of
Lucas' model of the size distribution of firms  \cite{Lucas}.


At the outset let us acknowledge a tension between our empirical
results and the theoretical model in this section. In our companion
paper and in the preceding section, we analyze the scaling properties
of the distribution of the logarithmic growth rate $r_1$ and its
standard deviation $\sigma_1$.  In this section we view companies as
consisting of many business units. Since the sales of a company are
the sum of the sales of individual units rather than their product, it
is more convenient to analyze the standard deviation of the annual
firm size change rather then the logarithmic growth rate.  Let
$\Sigma_1(S_0)$ be the standard deviation of end-of-period size for
initial size $S_0$. Since $\sigma_1 \sim {S_0}^{-\beta}$ and since
$S_1\equiv S_0 \exp(r_1) \approx S_0+S_0r_1$ , it follows that
$\Sigma_1(S_0) \approx S_0\sigma_1 \sim {S_0}^{1-\beta}$. As discussed
in Appendix A, $\sigma_1$ must be small for this approximation to
hold.

\subsection{Definition of the model}

Let us start by assuming that every company, regardless of its size, is
made up of similarly sized units.  Thus, a company of size $S_0$ is on
average made up of $N = S_0 / \bar{\xi}$ units, where
\begin{equation}
\bar{\xi} = \frac{1}{N} \sum_{i=1}^N \xi_i,
\label{e-units}
\end{equation}
and $\xi_i$ is the size of unit $i$.  We further assume that the
annual size change $\delta_i$ of each unit follows a bounded distribution
with zero mean and variance $\Delta$, which is independent of $S_0$.
It is important to notice that throughout this section and the
following we consider $\Delta \ll \bar{\xi}^2$, to insure that 
sizes  of  units remain positive. Since  some divisions after several cycles
of growth may shrink almost to zero, while others grow
several times, we assume that companies dynamically reorganize themselves 
so that they begin each period with approximately equal-sized divisions 
and the inequality $\Delta \ll \bar{\xi}^2$ holds.

If the annual size changes of the different units are
independent, then the model is trivial. Using the fact that $\langle
\delta_i \rangle = 0$, we have
\begin{equation}
\left \langle S_1 \right \rangle = 
S_0 +\sum_{i=1}^N \langle \delta_i \rangle = S_0.
\label{e-sigma-ind1}
\end{equation}
The second moment of the distribution is given by
\begin{eqnarray}
\left \langle {S_1}^2 \right \rangle & = & \left \langle \left ( 
         S_0 + \sum_{i=1}^N  \delta_i \right )^2 \right \rangle 
   =  {S_0}^2 +  \sum_{i=1}^N \sum_{j=1}^N \langle \delta_i \delta_j
\rangle  \\
\label{e-sigma-ind2}
\nonumber 
  & = & {S_0}^2 + N \Delta,
\end{eqnarray}
where we used again the fact that the $\delta_i$'s are centered and 
independent.

Thus, the variance in the size of the company is
\begin{equation}
{\Sigma_1}^2(S_0) = N \Delta = S_0 \frac{\Delta}{\bar{\xi}} \sim S_0.
\label{e-sigma-ind}
\end{equation}
Using the fact that $\Sigma(S_0) \sim {S_0}^{1-\beta}$ (see Appendix A),
it follows that $\beta=1/2$.

\begin{figure}
\narrowtext
\centerline{
\epsfysize=0.8\columnwidth{\rotate[r]{\epsfbox{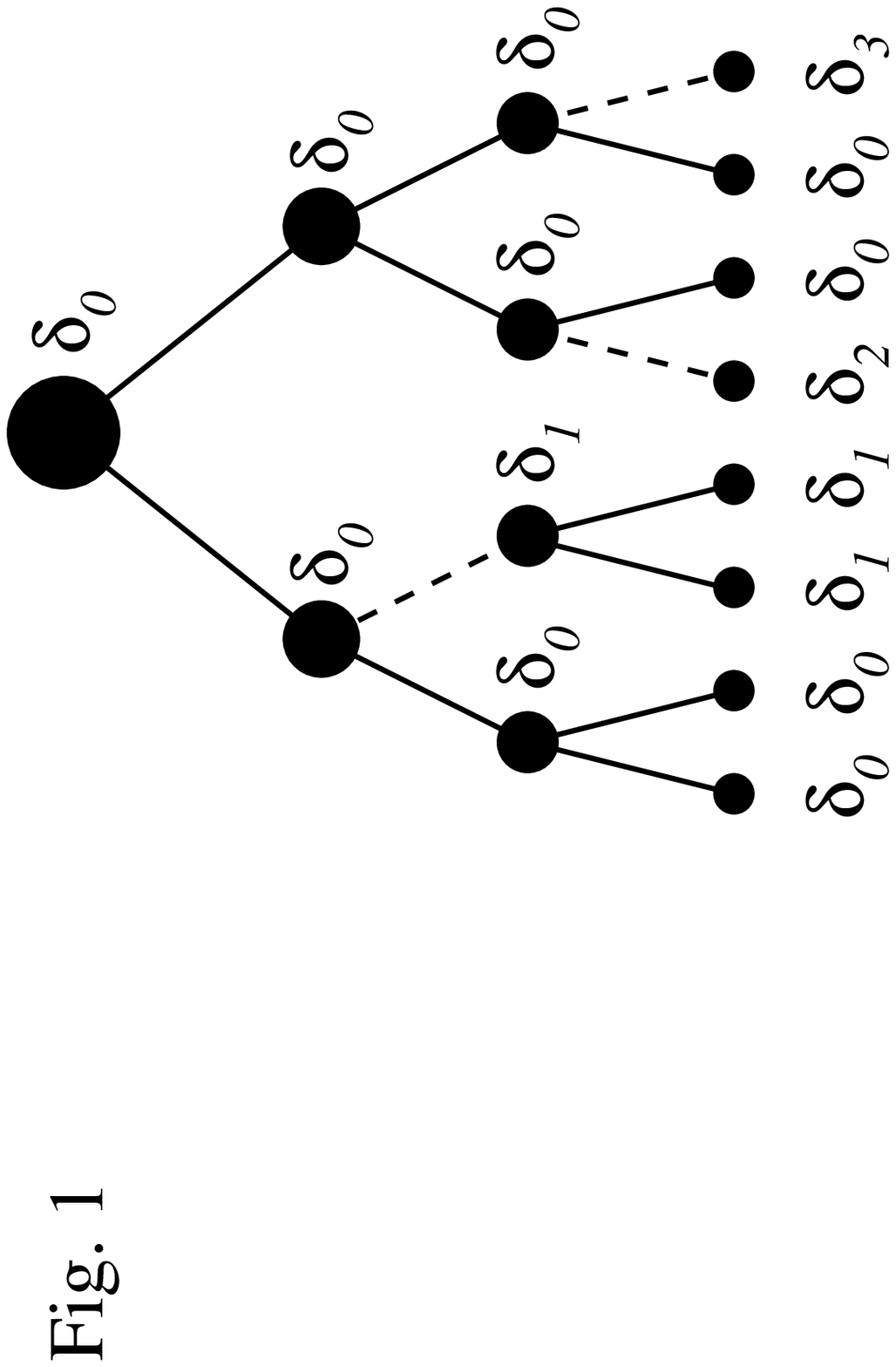}}}
}
\vspace*{1.0cm}
\caption{ The hierarchical-tree model of a company.  We represent a
        company as a branching tree.  Here, the head of the company
        makes a decision about the change $\delta_0$ in the size of
        the lowest level units.  That decision is propagated through
        the tree.  However, the decision is only followed with a
        probability $\Pi$.  This is pictured in the figure as a full
        link.  With probability $(1-\Pi)$ a new growth rate is
        defined.  This is pictured as a slashed link.  We see that at
        the lowest level there are clusters of values $\delta_i$ for
        the changes in size.  }
\label{f-tree}
\end{figure}

The much smaller value of $\beta$ that we find indicates the presence
of strong positive correlations among a company's units.  We can
understand this result by considering the tree-like hierarchical
organization of a typical company \cite{Radner}.  The head of the tree
represents the head of the company, whose policy is passed to the
level beneath, and so on, until finally the units in the lowest level
take action.  These units have again a mean size of
$\bar{\xi}=S_0/N$ and annual size changes with zero mean and variance
of $\Delta$.  Here we assume for simplicity that at every level other
than the lowest each node is connected to exactly $z$ units in the
next lowest level. Then the number of units $N$ is equal to $z^n$,
where $n$ is the number of levels (see Fig.~\ref{f-tree}).  


What are the consequences of this simple model?  Let us first assume
that the head of the company suggests a policy that could result in
changing the size of each unit in the lowest level by an amount
$\delta_0$.  If this policy is propagated through the hierarchy without
any modifications, then it is the same as assuming in
Eq.~(\ref{e-sigma-ind2}) that all the $\delta_i$'s are identical.  This
implies that
\begin{equation}
\left \langle {S_1}^2 \right \rangle =
      {S_0}^2 + N^2 \Delta,
\label{e-sigma-obey1}
\end{equation}
from which follows
\begin{equation}
{\Sigma_1}^2(S_0) = N^2 \Delta= S_0^2 \frac{\Delta}{\bar{\xi}^2},
\label{e-sigma-obey}
\end{equation}
and we conclude that $\beta =0$.

Of course, it is not realistic to expect that all decisions in an
organization would be perfectly coordinated as if they were all
dictated by a single ``boss.''  Hierarchies might be specifically
designed to take advantage of information at different levels; and
mid-level managers might even be instructed to deviate from decisions
made at a higher level if they have information that strongly suggests
that an alternative decision would be superior.  Another possible
explanation for some independence in decision-making is organizational
failure, due either to poor communication or disobedience.

\subsection{Analytical calculations}

To model the intermediate case between $\beta = 0$ and $\beta = 1/2$,
let us assume that the head of a company makes a decision to change
the size of the units of a company by an amount $\delta_0$.  We also
assume that $\delta_0$, for the set of all companies, has zero mean
and variance $\Delta$.  Furthermore, we consider that each manager at
the nodes of the hierarchical tree follows his supervisor's policy
with a probability $\Pi$, while with probability $(1-\Pi)$ imposes a
new {\it independent\/} policy.  The latter case corresponds to the
manager acting as the head of a smaller company made up of the units
under his supervision.  Hence the size of the company becomes a random
variable with a standard deviation that can be computed either with
numerical simulations or using recursion relations among the levels of
the tree.

Since the calculations are somewhat involved, we include them in
Appendix B for the interested reader (see also
Refs.~\cite{Harris,Li}).  The main result is that the variance of the
fluctuations in a $n$-level hierarchical tree is given by
\begin{equation}
{\Sigma_1}^2(n) = \Delta~\left ( z^n \frac{1 - \Pi^2}{1 - z \Pi^2}
                - (z \Pi)^{2n} \frac{(z - 1) \Pi^2}{1 - z \Pi^2}
                \right ).
\label{e.15x}
\end{equation}

If $z \Pi^2 > 1$, then $(z \Pi)^{2n}$ dominates the growth, and we get
\begin{equation}
{\Sigma_1}^2(n) \sim (z \Pi)^{2n} \sim N^2~{\Pi}^{2 \ln N / \ln z} 
        \sim N^2~N^{2 \ln \Pi / \ln z}
        \sim {S_0}^{2 + 2 \ln \Pi / \ln z},
\end{equation}
which implies $\beta = -\ln \Pi / \ln z$.  On the other hand, if $z
\Pi^2 < 1$, then $z^n=N$ is the dominant term, and we obtain
\begin{equation}
{\Sigma_1}^2(n) \sim z^n \sim N \sim S_0,
\end{equation}
which implies $\beta = 1/2$.

  Finally, we can write, for $n \gg 1$, that the hierarchical model
leads to
\begin{equation}
\beta = \cases{ -\ln \Pi / \ln z & if $\Pi > z^{-1/2}$, \cr
                    ~~~~1/2      & if $\Pi < z^{-1/2}$  \cr }
\label{e-model2}
\end{equation}
Even for small $n$, we find that Eq.~(\ref{e-model2}) is a good
approximation --- e.g., while for $z=2$ and $\Pi = 0.87$ we predict
$\beta = 0.20$, when we take $n=3$ the deviation from the predicted
value is only $0.03$, i.e., about $15\%$.

\begin{figure}
\narrowtext
\centerline{
\epsfysize=0.8\columnwidth{\rotate[r]{\epsfbox{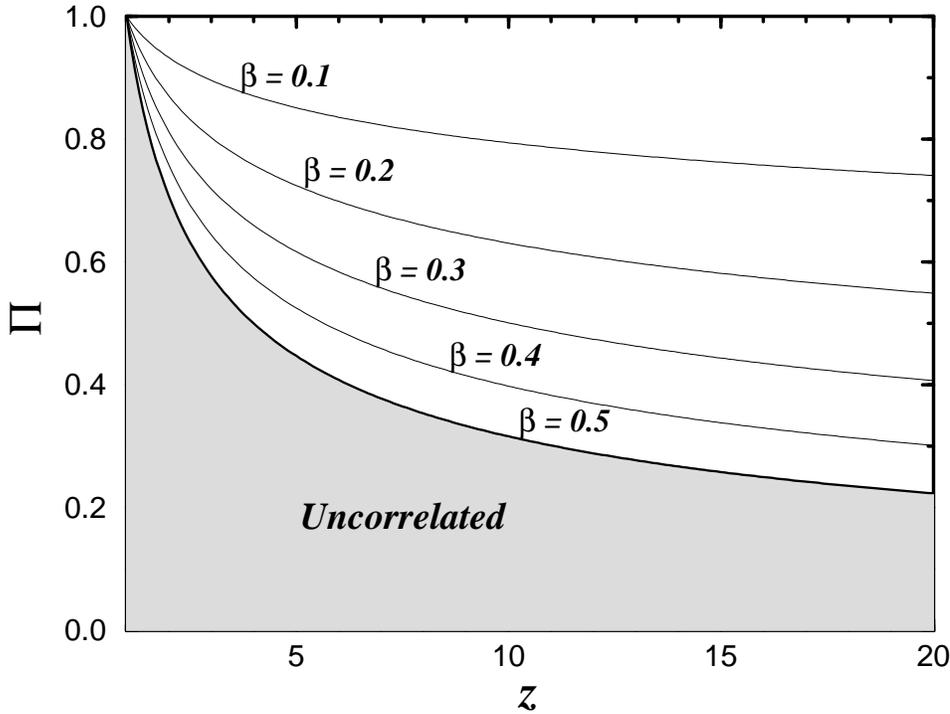}}}
}
\vspace*{1.0cm}
\caption{
        Phase diagram of the hierarchical-tree model.  To each pair of
values of $(\Pi,z)$ corresponds a value of $\beta$.  We plot the
iso-curves corresponding to several values of $\beta$.  In the shaded
area, marked ``Uncorrelated,'' the model predicts that $\beta = 1/2$,
i.e., that the units of the company are uncorrelated.  Our empirical
data suggests that most companies have values of $\Pi$ and $z$ in
close to the curve for $\beta = 0.2$.
}
\label{f-phase}
\end{figure}

Equation (\ref{e-model2}) is confirmed in the two limiting cases: when
$\Pi=1$ (absolute control) $\beta = 0$, while for all $\Pi <
1/z^{1/2}$, decisions at the upper levels of management have no
statistical effect on decisions made at lower levels, and $\beta =
1/2$.  Moreover, for a given value of $\beta < 1/2$ the control level
$\Pi$ will be a decreasing function of $z$: $\Pi = z^{-\beta}$,
cf. Fig.~\ref{f-phase}.  For example, if we choose the empirical value
$\beta\approx 0.15$, then Eq.  (\ref{e-model2}) predicts the plausible
result $0.9 \ge \Pi \ge 0.7$ for a range of $z$ in the interval $2 \le
z \le 10$.

\section{Combining the Two Models}

We started with two central empirical findings about firm growth rates.
The model in Section II predicts one of those findings (the shape of the
distribution) and the model in Section III predicts the other (the power
law dependence of the standard deviation of output on firm size).  This
section addresses the relationship between the two models.  First, we
address concerns that the models might be contradictory and show that
they are not.  Then, we show how the models can be combined into a
single model that predicts both of our empirical findings.
                
In the tree model, firm growth rates are potentially the result of many
independent decisions.  As a result, one might expect that the Central
Limit Theorem would imply a Gaussian distribution of firm output. In
fact, however, the distribution of outputs is not necessarily Gaussian.
                
To address the distribution of firm output in the tree model, it is
necessary to make an assumption about the distribution from which each
independent growth decision is drawn.  No such assumption is needed to
analyze the standard deviation of firm growth rates, but is needed to
analyze the shape of the distribution.
                
In Fig. 3, we show the distribution of the inputs (i.e., of each
independent decision) and the outputs for a tree with $z = 2$, $\Pi =
0.87$, and $n = 10$.  We find that for Gaussian distributed inputs,
the output is not Gaussian in the tails.  This finding is remarkable.
First of all, with $z = 2$ and $n = 10$, the firm consists of 1024
units.  With a  probability to disobey of $1 - 0.87 = 0.13$, one would
expect $0.13\times 1024\approx 133$ of the units to, on average, make
independent decisions about their growth rates. Thus, even for
non-Gaussian inputs, one can hypothesize that the output is close to
Gaussian. Moreover, for Gaussian inputs, the sum of independent
Gaussians is itself Gaussian. Thus, for every particular configuration of
the disobeying links, the output distribution is Gaussian with
variance $m\Delta$, which is a function of this random
configuration. However, there are $2^{(z^{n+1}-z)/(z-1)}$ possible
configurations of links each of which produce a Gaussian distribution
with different integer $m$.

\begin{equation}
p_n(S_1)=\sum_m p_m^n{1\over\sqrt{2\pi m\Delta}}
e^{-(S_1-S_0)^2/2m\Delta}, 
\label{e.1x}
\end{equation}
which is no longer Gaussian for the observed form of $p_m^n$.

\begin{figure}
\narrowtext
\centerline{
\epsfysize=0.8\columnwidth{\rotate[r]{\epsfbox{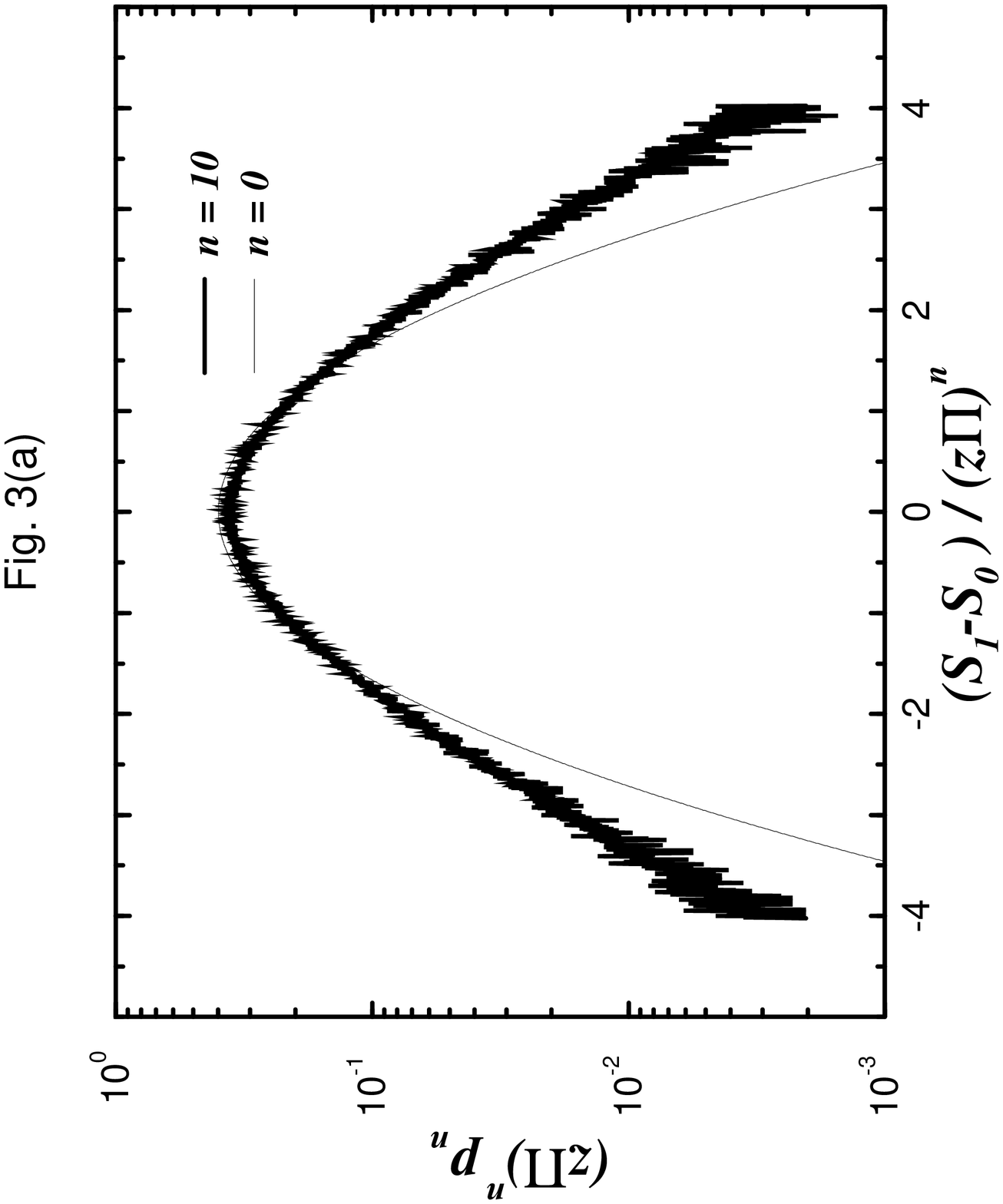}}}
}
\vspace*{1.0cm}
\centerline{
\epsfysize=0.8\columnwidth{\rotate[r]{\epsfbox{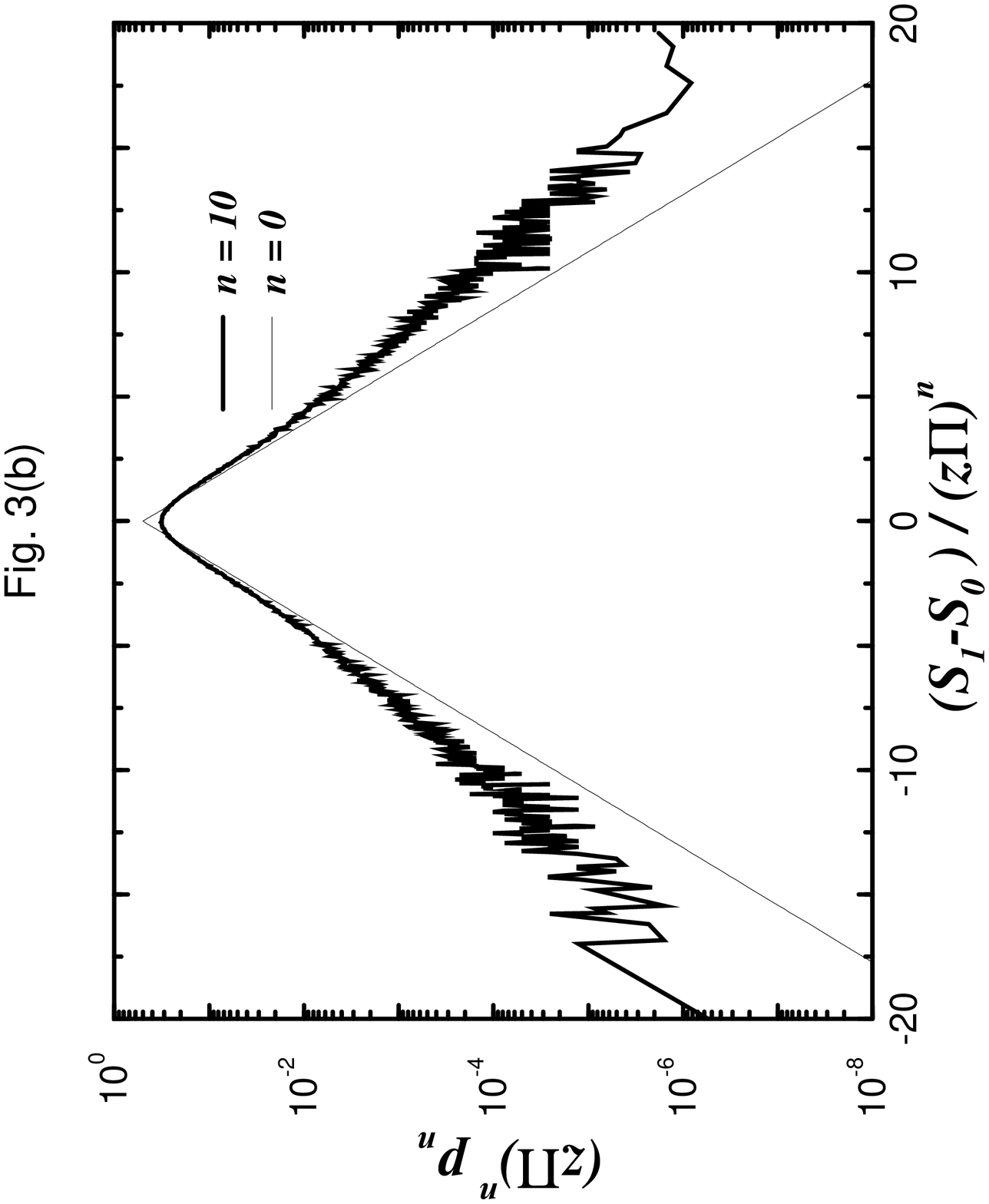}}}
}
\vfill
\centerline{
\epsfysize=0.8\columnwidth{\rotate[r]{\epsfbox{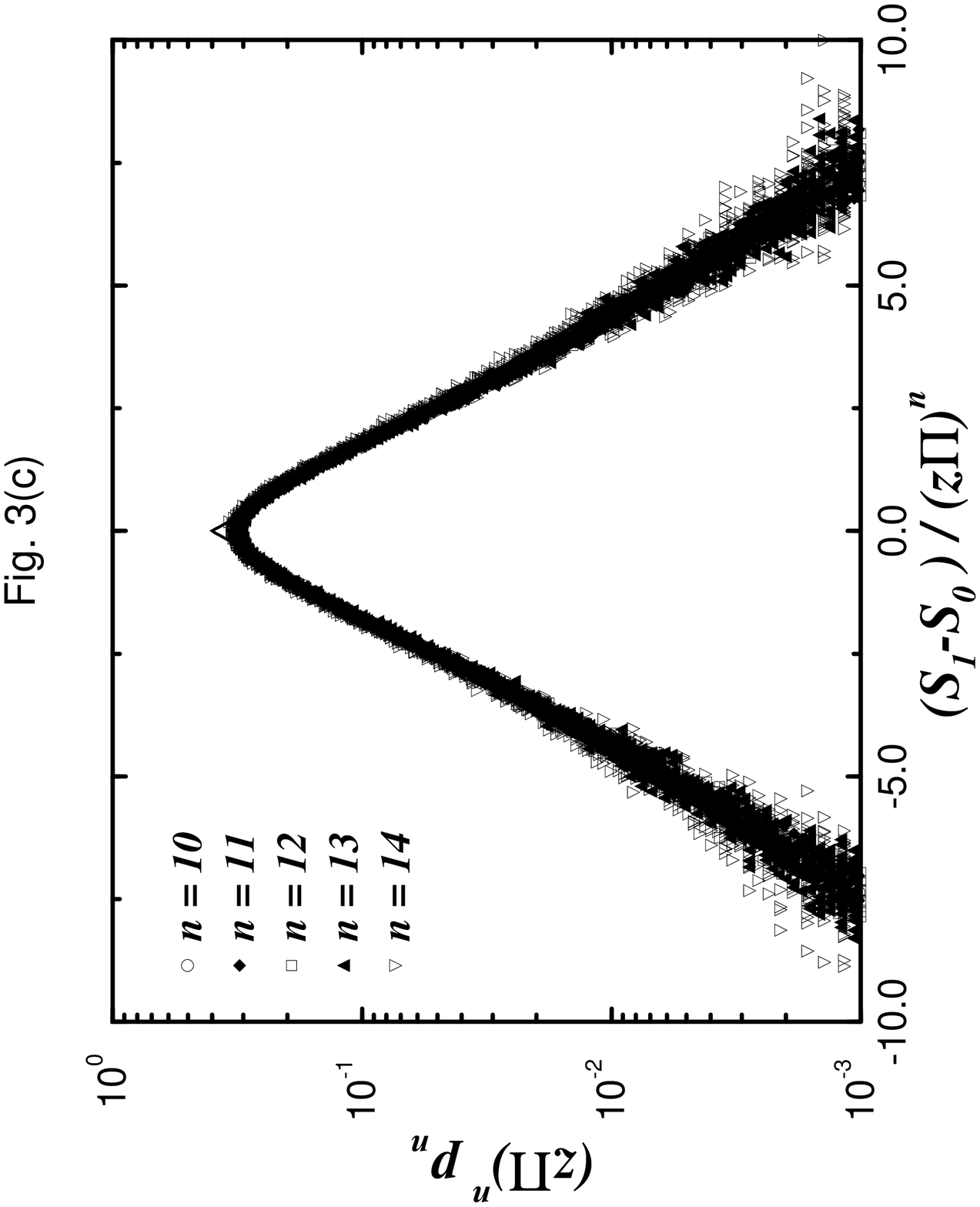}}}
}
\vspace*{1.0cm}
\caption{
        Probability density for the output and input variables in the
tree model.  Here we have $z = 2$, $\Pi = 0.87$, and $n = 10$.
        (a) Gaussian distribution of the input.  
        (b) Exponential distribution of the input.
        (c) Data collapse of the output distribution for trees with
different number of levels $n$.  The other parameters remain unchanged 
and the input is exponentially distributed.  It is visually apparent 
the similarity of the numerical results with the empirical data of 
Fig.~4(b) \protect\cite{Empirical}.}
\label{f-output}
\end{figure}

 Figure 4 shows the probability $p_m^n$ to get a tree with given $m$
computed for all trees with a given number of levels $n$, $\Pi=0.87$,
and $z=2$. As visually apparent in Fig.~4, this probability density is
a non-trivial function, which is discussed in more detail in Appendix
C. The final distribution of the firm output $S_1$ will be thus given
by the convolution of two densities: $p_m^n$ and Gaussian with
variance $m\Delta$

In a general case, it can be shown by martingale theory \cite{Feller}
that for any input distribution $f(x)$ with zero mean and finite
variance $\Delta$, the output distribution converges for $n\to\infty$ to
a distribution
\begin{equation}
{1\over\sum_1(n)}g_f\left({x\over\sum_1(n)}\right),
\label{e.2x}
\end{equation}
where $g_f$ is a function that does not depend on $n$ but depends on
$f$.  Thus, we cannot expect to obtain a result that the output
distribution must be exponential regardless of the input distribution.
It would, however, be desirable to find some simple input distribution
that yields the output distribution that we actually observe.  Figure
3 also shows the output distribution when the input distribution is
exponential in terms of $S_1-S_0$. For small $\sigma_1$, it
practically coincides with Eq.~(\ref{model12}).  In this case, the
output distribution is nearly exponential, and the slightly fatter
wings that we observe are arguably consistent with our empirical
results.  Thus, in the limit of small $\sigma_1$, we can combine the
models of the two sections by assuming that the dynamic process
described in Sect. II provides the input distribution for the tree
model in Sect. III.  This additional assumption in the tree model then
predicts both of our empirical findings. For large $\sigma_1$, the
direct combination of two models needs additional fine-tuning.

\begin{figure}
\narrowtext
\centerline{
\epsfysize=0.8\columnwidth{\rotate[r]{\epsfbox{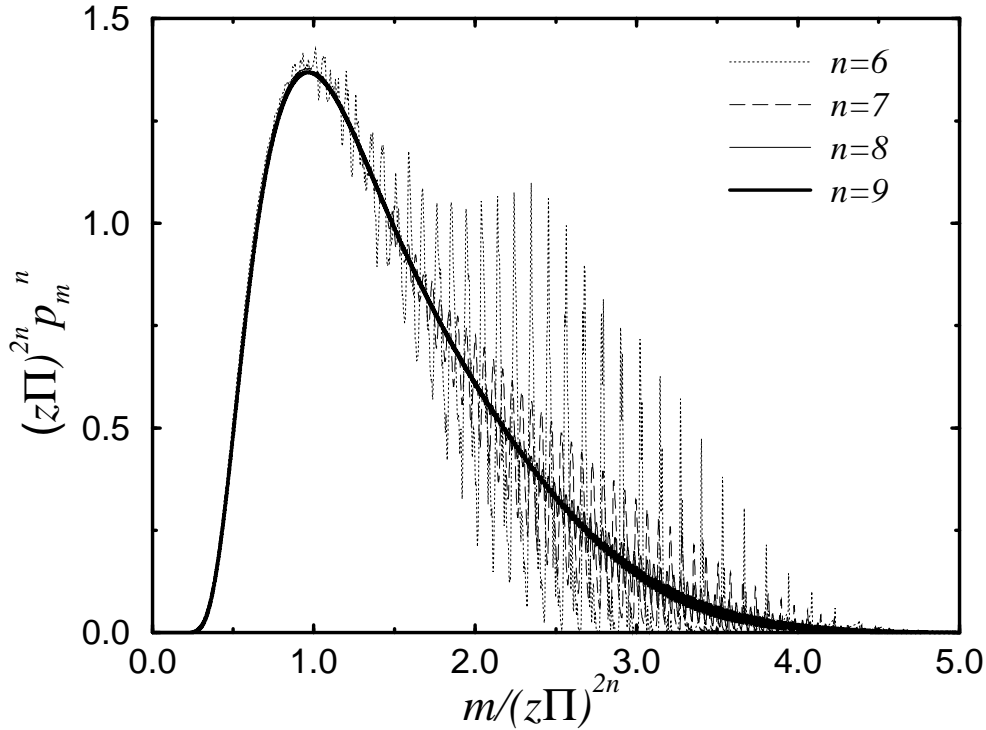}}}
}
\vspace*{1.0cm}
\caption{ 
        Numerical estimation from exact enumeration of the coefficients
$p_n^m$ of the generating function $p_n(s)$.  It is visually apparent
that the coefficients scale according to Eq.~(B20) even for $n$ as small
as 6.  This result suggests that the companies have a self-similar 
structure.}
\label{f-coef}
\end{figure}

\section{Conclusions}
                
The two central results of our previous paper are that the distribution
of company growth rates is exponential and the standard deviation of
growth rates scales as a power law of firm size with scaling exponent 
$\beta \approx -0.2$.  Any realistic theory of the firm in economics must be
consistent with these empirical findings.  In this paper, we have
presented simple models that are consistent with our empirical findings.
Indeed, the models have only two key assumptions.  One is that each
company has a natural size and the other is that decisions in
hierarchical organizations are positively but imperfectly correlated.
These models suggest that very simple mechanisms may provide insight
into our empirical findings.

One limitation of the model in this paper is that it only predicts our
results about one year growth rates.  A complete model of the firm would
also predict the distribution of growth rates over longer horizons.  We
believe that extending this model to additional periods would not
provide a complete description of firm dynamics.  In reality, the
standard deviation of growth rates goes up as the time horizon
increases.  The attraction in our model to a stable company size
prevents the distribution from spreading over time as much as we
actually observe.

\acknowledgments

We thank R. N. Mantegna for important help in the early stages of
this work, and JNICT (L.A.), DFG (H.L. and P.M.), and NSF for
financial support.

\appendix

\section{Width of the Distribution of Final Sizes}

The theory in Section III establishes results about the standard
deviation of the growth rate.  Our empirical results in the earlier
paper concerned the standard deviation of the logarithmic growth rate
defined as $\ln(S_1/S_0)$.  This appendix establishes the exact
relationship between the standard deviation of the growth rate and the
standard deviation of the logarithmic growth rate.  Thus we will compute
the width of the distribution of final sizes $S_1 \equiv S_0 \exp{r_1}$,
that we designate by $\Sigma_1(S_0)$.  We can express $\Sigma_1$ as
\begin{equation}
\Sigma_1(S_0)^2 = \left \langle {S_1}^2 \right \rangle -
\left \langle {S_1} \right \rangle^2.
\label{e-s1}
\end{equation}
Taking $\bar{r_1}(s_0) \approx 0$, and assuming that the standard
deviation of the distribution is small ($\sigma_1 < 1/\sqrt{2}$ 
which holds for companies with sales larger than $10^6$ dollars,)
simple integrations lead to
\begin{equation}
\left \langle S_1 \right \rangle = \int_{-\infty}^{+\infty} S_1~
p(r_1|S_0)~dr_1 = \frac{S_0}{1 - {\sigma_1}^2 / 2},
\end{equation}
and
\begin{equation}
\left \langle {S_1}^2 \right \rangle = \int_{-\infty}^{+\infty}
{S_1}^2~p(r_1|S_0)~dr_1 = \frac{{S_0}^2}{1 - 2 {\sigma_1}^2}.
\end{equation}
Replacing these results onto (\ref{e-s1}) and expanding in Taylor
series, we obtain
\begin{eqnarray}
\Sigma_1(S_0)^2 & = & {S_0}^2 \left( 1 + 2{\sigma_1}^2 + 4{\sigma_1}^4 +
\cdots - 1 - {\sigma_1}^2 - 3{\sigma_1}^4 / 4 + \cdots \right) \nonumber
\\
          & \approx & (S_0 \sigma_1)^2 (1 + 13{\sigma_1}^2 / 4).
\end{eqnarray}
Thus, to first order, we obtain
\begin{equation}
\Sigma_1(S_0) \sim {S_0}^{1-\beta}.
\label{e-sigma2}
\end{equation}

\section{Analytical calculation of the variance of the growth rate 
for the hierarchical-tree model}

This appendix provides a rigorous derivation of Eq.~(\ref{e.15x})
Let, as before, $S_1$ represent the final size of a company with
initial size $S_0$, and assume that the company has $n$ levels in its
hierarchical tree.  According to the rules of the model, the decision
of the head of the company will only be followed by those units in the
bottom level which are connected to the top by a chain of managers with
``obeying links.''  Thus, the number of units of the company that
follow the policy of the head of the company $T_n$ can be related to
the well known problem of the number of male descendents of a family
after $n$ generations \cite{Harris}. The solution is that for a
$n$-level tree with $z$ branches the average number of units at the
end is given by 

\begin{equation}
\langle T_n \rangle = (z \Pi)^n.
\label{e-connected}
\end{equation}


Now, let us look at the problem of calculating $\Sigma_1(S_0)$. Our
problem is slightly more complicated since it includes double
averaging over all realizations of growth rates of independent units
and over all possible configurations of the tree. Let us look at the
$n^{\mbox{\scriptsize th}}$ level of a tree with a certain
configuration of obeying and disobeying links. We can define clusters
of units connected to one another through obeying links. Let us suppose
that there are $M_n$ distinct clusters of size $\nu_i$.
According to the rules of the model, all units in cluster $i$ share
the same value of the annual change $\delta_i$. Thus, he final size of
the company will be
\begin{equation}
S_1 = S_0+ \sum_{i=1}^{M_n} \nu_i \delta_i,
\end{equation}
where $\delta_i$ are independent random variables with zero mean and
variance $\Delta$.

The variance in $S_1$, for a {\it given\/} tree with $n$ levels, can
be obtained by averaging over all realizations of $\delta_i$,
\begin{equation}
\Delta_n  =  \Delta \sum_{i=1}^{M_n} \nu_i^2  \equiv  m_n \Delta,
\end{equation}
where $m_n$ is a random variable depending solely on the structure of
the tree
\begin{equation}
m_n = \sum_{i=1}^{M_n} \nu_i^2.
\end{equation}


To obtain ${\Sigma_1}^2$ we need now to average over all possible
configurations of the hierarchical tree
\begin{equation}
{\Sigma_1(n)}^2 = \Delta \langle m_n \rangle.
\label{e-step}
\end{equation}
In order to calculate $\langle m_n \rangle$, we will start by
computing the conditional average value $\left.\langle
  m_n\rangle\right|_{m_{n-1}}$, where $m_{n-1}$ refers to the previous
level on the tree.  A cluster of size $\nu_i$ in the $(n-1)$ level is
connected to $z \nu_i$ units in the $n$-level; $\nu_i'$ of the links
are obeying, while $(z \nu_i - \nu_i')$ are disobeying. The obeying
links will give rise to a cluster of size $\nu_i'$ in level $n$, while
the disobeying links give rise to $(z \nu_i - \nu_i')$ clusters of
size one.  Thus, we have
\begin{eqnarray}
m_n & = & \sum_{i=1}^{M_{n-1}} \left( {\nu_i'}^2 + 
  (z \nu_i - \nu_i') \right) \nonumber \\
    & = & \sum_{i=1}^{M_{n-1}} ({\nu_i'}^2 - \nu_i') + 
    z \sum_{i=1}^{M_{n-1}} \nu_i \nonumber \\
    & = & \sum_{i=1}^{M_{n-1}} ({\nu_i'}^2 - \nu_i') + z^n.
\end{eqnarray}
The probability of a configuration with a $\nu_i'$ obeying links is
\begin{equation}
{z \nu_i \choose \nu_i'} \Pi^{\nu_i'} (1 - \Pi)^{z \nu_i - \nu_i'}.
\end{equation}
By averaging over all possible configurations of links, we obtain
\begin{equation}
\langle m_n \rangle |_{m_{n-1}} = \sum_{i=1}^{M_{n-1}} \left\{
  \sum_{\nu_i'=0}^{z \nu_i} {z \nu_i \choose \nu_i'}
  \Pi^{\nu_i'} (1 - \Pi)^{z \nu_i - \nu_i'} ({\nu_i'}^2 - \nu_i')
  \right\} + z^n.
\label{e-x77}
\end{equation}
The series in (\ref{e-x77}) can be calculated with one of the
traditional ``tricks.'' Defining $q=1-\Pi$, $k=z\nu$, and $j=\nu_i'$,
we have
\begin{eqnarray}
\sum_{j=0}^k {k \choose j} (j^2 - j) \Pi^j (1 - \Pi)^{k-j} & = &
\Pi^2 {\partial^2 \over \partial \Pi^2} (\Pi+q)^s |_{\Pi+q=1} \nonumber \\
& = & k (k - 1) \Pi^2.
\end{eqnarray}
Replacing this result into (\ref{e-x77}), we obtain
\begin{eqnarray}
\langle m_k \rangle |_{m_{n-1}} & = & (z \Pi)^2 
\sum_{i=1}^{M_{n-1}} \nu_i^2 - \Pi^2 z \sum_{i=1}^{M_{n-1}} \nu_i 
+ z^n \nonumber \\
& = & (z \Pi)^2 m_{n-1} + (1 - \Pi^2) z^n.
\label{e.xx}
\end{eqnarray}

Hence $\langle m_n \rangle$ satisfies the recursion relation
\begin{equation}
\langle m_n \rangle = (z\Pi)^2 \langle m_{n-1} \rangle + (1 - \Pi^2)z^n,
\qquad\quad \langle m_0 \rangle = 1.
\label{e.xxx}
\end{equation}
Writing the first few terms in the succession and induction show that
\begin{equation}
\langle m_n \rangle = (z\Pi)^2 + (1 - \Pi^2) z^n 
\sum_{i=0}^{n-1} (z \Pi^2)^i.
\end{equation}
Replacing the geometric series by its value and simple calculations
lead to
\begin{equation}
\langle m_n \rangle = \left ( z^n \frac{1 - \Pi^2}{1 - z \Pi^2}
                - (z \Pi)^{2n} \frac{(z - 1) \Pi^2}{1 - z \Pi^2}
                \right ).
\end{equation}
Replacing this result into Eq.~(\ref{e-step}), we get
\begin{equation}
\Sigma_1(n)^2 = \Delta ~\left ( z^n \frac{1 - \Pi^2}{1 - z \Pi^2}
                - (z \Pi)^{2n} \frac{(z - 1) \Pi^2}{1 - z \Pi^2}
                \right ).
\end{equation}

\section{Distribution of the output variable for the 
hierarchical-tree model}

In this appendix we will derive the dependence of the variance of the
distribution of growth rates for the hierarchical-tree model in a more
formal way.  At the same time we will get some insight onto the
distribution of the number of end units that are connected by obeying
links to the head of the tree.  We will concentrate on the case in
which the distribution of inputs is Gaussian.

Let us look at the the $n$-th level of the tree: We can define
clusters of units which are connected to one another, in the tree,
through obeying links.  Thus, they share the same value of the annual
size change.  Supposing there are $M_n$ distinct clusters with sizes
$\nu_i$, we have
\begin{equation}
N = z^n = \sum_{i=1}^{M_n} \nu_i.
\end{equation}
Since there is a set of possible tree structures for any given value
of $\Pi$ (and $n$ and $z$), we should consider the set of all possible
values of $M_n$ ($1 \leq M_n \leq N$).  Let us then denote the set of
all partitions of $N$ into different clusters as $\Phi$, and each of
these partitions as $\phi_i$.  Naturally, the sum of the probabilities
of each partition $P(\phi_i)$ verifies
\begin{equation}
1 = \sum_{\phi_i \epsilon \Phi} P(\phi_i).
\end{equation}
It is known that for {\it large\/} values of $N$, the number of
different partitions behaves as $1/(\sqrt{48}N)~\exp{(\pi
  \sqrt{N/3})}$ \cite{Handbook}.

Let us denote the probability density of the input variable $\delta$
as $f(\delta)$.  The probability density for the output of a cluster
of $s$ units connected by obeying links is
\begin{equation}
f(x = s \delta) = \frac{1}{s}~ f \left( \frac{x}{s} \right).
\end{equation}
Thus, the distribution of the output variable $S = \sum_j^{M_i} x_j$
is given by
\begin{equation}
p_n(S) = \sum_{\phi_i \epsilon \Phi} 
    \frac{1}{s_1}~ f \left ( \frac{x_1}{s_1} \right ) ~\ast~
    \frac{1}{s_2}~ f \left ( \frac{x_2}{s_2} \right ) ~\ast~ ... 
     ~\ast~ \frac{1}{s_{M_i}}~ f \left ( 
     \frac{x_{M_i}}{s_{M_i}} \right ) P(\phi_i).
\label{e-prob-s}
\end{equation}
where $g(y = x_1+x_2) = f(x_1)~\ast~f(x_2) = \int f(x_1) f(y-x_1) dx_1$.


If $\delta$ is assumed to be Gaussian distributed with zero mean and
unit variance:
\begin{equation}
f(\delta) = \frac{1}{\sqrt{2\pi}} \exp \left( 
  -\frac{\delta^2}{2} \right),
\end{equation}
then the convolution leads to
\begin{eqnarray}
g(x_1+x_2) & = & f(x_1)~\ast~f(x_2) \nonumber \\
           & = & \frac{1}{\sqrt{2\pi(s_1^2+s_2^2)}}~
      \exp \left( -\frac{y^2}{2(s_1^2+s_2^2)} \right)
      \nonumber \\
           & = & \frac{1}{\sqrt{s_1^2 + s_2^2}}~ f \left ( 
      \frac{y}{\sqrt{s_1^2 + s_2^2}} \right ).
\end{eqnarray}
Replacing this result onto (\ref{e-prob-s}), we obtain
\begin{equation}
p_n(S) = \sum_{\phi_i \epsilon \Phi} 
\frac{1}{\sqrt{\sum_{j=1}^{M_i} s_i^2 }}~ f \left( 
\frac{S^2}{\sqrt{\sum_{j=1}^{M_i} s_i^2}} \right ).
\label{e-prob-s2}
\end{equation}
A simple analysis of Eq.~(\ref{e-prob-s2}) shows that any two
partitions, $\phi_i$ and $\phi_k$, are equivalent in terms of their
output distributions if they verify
\begin{equation}
\sum_{j=1}^{M_i} s_j^2 = \sum_{j=1}^{M_k} s_j^2 = m.
\label{e-prob-eq}
\end{equation}
On the other hand, the triangular (or Schwarz) inequality allows us to
determine the possible number of partitions that are not equivalent
because of the constraints on the value of $s$
\begin{equation}
N  = \sum_{j=1}^{N} 1^2 \le \sum_{j=1}^{M_i} s_i^2 \le 
\left ( \sum_{j=1}^{M_i} s_i \right ) ^2 = N^2.
\label{e-prob-number}
\end{equation}

Equations (\ref{e-prob-eq}-\ref{e-prob-number}) imply that the sum in
(\ref{e-prob-s2}) over different partitions can be replaced by a sum
over $m$.  Thus, we can write asymptotically
\begin{equation}
p_n(S) = \sum_{m=N}^{N^2} \frac{p_m^n}{\sqrt{m}}~ f \left( 
\frac{S}{\sqrt{m}} \right),
\end{equation}
and, finally,
\begin{equation}
p_n(S) = \sum_{m=N}^{N^2} \frac{p_m^n}{\sqrt{2 \pi m}}~ 
e^{- S^2 / 2m},
\label{e-prob-out}
\end{equation}
where $p_m^n$ is the total probability of all equivalent partitions with
given $n$ and $m$.

The standard way to calculate the coefficients $p_m$ is to introduce a
generating function \cite{Harris}
\begin{equation}
p_n(S) = \sum_{m=N}^{N^2} p_m^n S^m,
\label{e-prob-out2}
\end{equation}
which is a polynomial of order $N^2$. To obtain the recursion
relations for $p_n(S)$, we need to distinguish the cluster of units
which is connected to the top of the tree from those clusters that are
not.  For each level $n$ we have a matrix of coefficients $p_{\ell,k}$
that characterizes the probability of the partition with the cluster
of $\ell$ elements connected to the head of the tree and the sum of
squares of the rest of the cluster sizes equal to $k$.  Thus, we can
look at the tree as made of two parts, the one connected to the top,
with size $\ell$, and the remaining of size $(N - \ell)$.  Here we
introduce the full generating function
\begin{equation}
p_n(y,S) = \sum_{\ell=0}^{N}~ \sum_{k=N-\ell}^{(N-\ell)^2} 
p_{\ell,k}^n~ y^{\ell}~ S^k,
\label{e.a13}
\end{equation}
where $m = {\ell}^2 + k$.

The reduced generating function $p_n(S)$ can be obtained from the full
generating function $p_n(y,S)$ if one formally puts
$y^\ell=S^{\ell^2}$ in Eq.~(\ref{e.a13}). In order to obtain the
recursion relation for the full generating function, let us consider a
tree with $n+1$ levels as $z$ trees connected by another level of
branches to the top. If a $n-$level tree is connected to the top by a
disobeying link, which happens with probability $(1-\Pi)$, its
clusters are totally independent of the other branches and we can use
the reduced generating function $p_n(S)$. If, however, a $n-$level
tree is connected to the top by a obeying link, which happens with
probability $\Pi$, its clusters merge with the clusters of other such
trees, and the full generating function $p_n(y,S)$ must be used.  Thus,
the generating function of level $n+1$ is related to the generating
function of level $n$ through the recursion relation
\begin{equation}
p_{n+1}(y,S) = \left( \Pi p_n(y,S) + (1 - \Pi) p_n(S) \right)^z.
\label{e-recursion}
\end{equation}
Unfortunately, to our knowledge, this recursion relation is too
complex to allow any simplification or solution.  Thus, we cannot
obtain the distributions of cluster sizes for the different values of
$n$.  On the other hand, the problem of obtaining the average value of
$\ell$ (which was earlier designated $\langle T_n \rangle$) and the
variance $\Sigma_1(n)^2$ of the output variable is relatively simple
\cite{Harris}.  Indeed
\begin{equation}
\langle T_n \rangle = \frac{\partial}{\partial y} 
                                  p_n(y,S)|_{y=1,S=1}.
\label{e-rec-tn}
\end{equation}
Combining (\ref{e-recursion}) and (\ref{e-rec-tn}), we obtain
\begin{eqnarray}
\langle T_{n+1}  \rangle & = & \frac{\partial}{\partial y}
           p_{n+1}(y,S)|_{y=1,S=1} \nonumber \\
                         & = & z \Pi~\frac{\partial}{\partial y}
           p_n(y,S)|_{y=1,S=1} \nonumber \\
                         & = & z \Pi~ \langle T_n \rangle.
\end{eqnarray}
And we recover Eq.~(\ref{e-connected}).  The variance can also be
easily obtained as \cite{Harris}
\begin{eqnarray}
\langle m_n\rangle & = &  \frac{\partial}{\partial S} p_n(S)|_{S=1}
                 \nonumber \\
                   & = &  \frac{\partial}{\partial y} y 
                 \frac{\partial}{\partial y} p_n(y,S)|_{y=1,S=1}
                       +  \frac{\partial}{\partial S}
                 p_n(y,S)|_{y=1,S=1},
\label{e-rec-sn}
\end{eqnarray}
which, after some algebra, leads to
\begin{equation}
\langle m_{n+1}\rangle = z\langle m_n\rangle + z(z-1) \Pi^2 
\langle T_n \rangle^2,
\end{equation}
which is equivalent to Eq.~(\ref{e.xxx}).

Although, as discussed earlier, the coefficients $p_m^n$ cannot be
calculated analytically, we can use Eq.~(\ref{e-recursion}) to find
their values numerically (see Fig.~4). Moreover, for $z \Pi^2>1$ the
coefficients $p_m^n$ of the reduced generating function $p_n(S)$ scale
as
\begin{equation}
p_m^n={1\over(z \Pi)^{2n}}~ g \left({m\over(z \Pi)^{2n}}\right)
\label{e.A20}
\end{equation}
for large $n$. This can be proven applying the martingale theory
\cite{Feller}.  Indeed, the sequence
\begin{equation}
\tilde m_n = \frac{m_n}{(z \Pi)^{2n}} + \frac{1-\Pi^2}{1 - z \Pi^2}
\left( 1 - \frac{1}{(z \Pi^2)^n} \right) 
\end{equation} 
obeys the martingale conditions: From Eq.~(\ref{e.xx}) it follows that
$\langle\tilde m_n\rangle|_{m_{n-1}}=\tilde m_{n-1}$. It also can be
shown that $\tilde m_n$ has limited variance for any $n$, and hence it
follows that for large $n$ the scaling relation (\ref{e.A20}) is
valid.

\end{document}